\documentclass[letterpaper]{article} 
\usepackage{latex/aaai24}  
\usepackage{times}  
\usepackage{helvet}  
\usepackage{courier}  
\usepackage[hyphens]{url}  
\usepackage{graphicx} 
\usepackage{natbib}  
\usepackage{caption} 
\usepackage{algorithm}
\usepackage[noend]{algorithmic}

\usepackage{newfloat}
\usepackage{listings}
\DeclareCaptionStyle{ruled}{labelfont=normalfont,labelsep=colon,strut=off} 
\floatstyle{ruled}
\newfloat{listing}{tb}{lst}{}
\floatname{listing}{Listing}
\title{Self-guided Knowledgeable Network of Thoughts: \\ Amplifying Reasoning with Large Language Models}
\author {
    Chao-Chi Chen\textsuperscript{\rm 1},
    Chin-Yuan Yeh\textsuperscript{\rm 1,3},
    Hsi-Wen Chen\textsuperscript{\rm 2,3},
    De-Nian Yang\textsuperscript{\rm 3},
    Ming-Syan Chen\textsuperscript{\rm 1,2},
}
\affiliations {
    \textsuperscript{\rm 1}Graduate Institute of Communication Engineering, National Taiwan University, Taiwan\\
    \textsuperscript{\rm 2}Department of Electrical Engineering, National Taiwan University, Taiwan\\
    \textsuperscript{\rm 3}Institute of Information Science, Academia Sinica, Taiwan\\
    \{chao-chi-chen, cyyeh, hwchen\}@arbor.ee.ntu.edu.tw, dnyang@iis.sinica.edu.tw, mschen@ntu.edu.tw
}

\usepackage{multirow}
\usepackage{fontawesome}
\usepackage{booktabs}
\usepackage{xspace}
\usepackage{tabularx}
\usepackage[dvipsnames]{xcolor}
\usepackage{tcolorbox}
\usepackage{amsfonts}
\usepackage{amsthm}
\newtheorem{definition}{Definition}

\newcommand{\n}[1]{#1}

\usepackage{amsmath}

\newcommand{\tagger}[1]{\textless#1\textgreater}
\newcommand{\mybox}[1]{\begin{tcolorbox}[colback=white, colframe=black, coltitle=black, boxrule=1pt, arc=0mm, left=1mm, right=1mm, top=1mm, bottom=1mm]{\textit{#1}}
\end{tcolorbox}}

\newcommand{\methodfull}{Knowledgeable Network of Thoughts\xspace}
\newcommand{\method}{kNoT\xspace}

\newcommand{\kprompt}{\textsl{Knowledge Extraction Prompt}\xspace}
\newcommand{\tprompt}{\textsl{\script Translation Prompt}\xspace}

\newcommand{\scriptfull}{LLM Workflow Template\xspace}
\newcommand{\script}{LWT\xspace}

\newcommand{\github}{\url{anonymous.4open.science/r/kNoT-5048}\xspace}

\newcommand{\1}{{\raisebox{.5pt}{\textcircled{\raisebox{-1pt} {1}}}}}
\newcommand{\2}{{\raisebox{.5pt}{\textcircled{\raisebox{-1pt} {2}}}}}
\newcommand{\3}{\raisebox{.5pt}{\textcircled{\raisebox{-1pt} {3}}}}
\newcommand{\4}{\raisebox{.5pt}{\textcircled{\raisebox{-1pt} {4}}}}
\newcommand{\5}{\raisebox{.5pt}{\textcircled{\raisebox{-1.1pt} {5}}}}
\newcommand{\6}{\raisebox{.5pt}{\textcircled{\raisebox{-1pt} {6}}}}

\begin{document}
\maketitle
\begin{abstract}
    We introduce \methodfull (\method): a prompt scheme that advances the capabilities of large language models (LLMs) beyond existing paradigms like Chain-of-Thought (CoT), Tree of Thoughts (ToT), and Graph of Thoughts (GoT). The key innovation of \method is the \textit{\scriptfull (\script)}, which allows for an executable plan to be specified by LLMs for LLMs. \script allows these plans to be \emph{arbitrary networks}, where single-step LLM operations are nodes, and edges correspond to message passing between these steps. Furthermore, \script supports selection of individual elements through indexing, facilitating \method to produce intricate plans where each LLM operation can be limited to elementary operations, greatly enhancing reliability over extended task sequences. We demonstrate that \method significantly outperforms the state of the art on six use cases, while reducing the need for extensive prompt engineering. 
    For instance, \method finds $92\%$ accuracy for sorting $32$ numbers over $12\%$ and $31\%$ for ToT and GoT, while utilizing up to $84.4\%$ and $87.3\%$ less task-specific prompts, respectively. 
\end{abstract}


\section{Introduction}

While instruction learning enables large language models (LLMs) to quickly adapt to new tasks using natural language prompts, performance significantly deteriorates in complex multi-step reasoning under zero-shot settings because the autoregressive token generation process does not align with the required reasoning sequence~\cite{mcleish2024transformers}.\footnote{Zero-shot denotes directly inferencing LLMs with a question.} To address this issue, research has designed transformers with specialized architectures to improve zero-shot performance for algorithmic reasoning~\cite{yang2023gpt, paul2023refiner}. However, these approaches are resource-intensive, requiring significant data curation and computational costs due to the use of multi-layered transformers~\cite{vaswani2017attention}. Moreover, the developed models are limited to specific problems, such as arithmetic operations~\cite{mcleish2024transformers} or symbolic tasks~\cite{gao2024efficient}, rather than serving as a generalized solution for multiple tasks. Additionally, striving for zero-shot capability in multi-step reasoning remains paradoxical, as even humans rely on intermediate reasoning steps for such tasks~\cite{saxton2019analysing}.

In contrast, \emph{prompt engineering} provides an economical approach by instructing LLMs with a set of \emph{prompts} to perform multi-step reasoning tasks~\cite{sahoo2024systematic}. For example, Chain-of-Thought (CoT)~\cite{wei2022chain} leverages \textit{intermediate thought processes} by instructing LLMs to ``think step by step,'' expecting them to follow an example solution with detailed intermediate steps as a guideline. However, CoT still operates with a single-inference approach, limiting its performance to support complex reasoning processes. Specifically, during a CoT operation, LLM needs to process all information (including the example, the input task, and previous steps) to generate the next \emph{thought}. As a result, the generated step-by-step solution often deviates from the provided example~\cite{stechly2024chain}, and the performance deteriorates as the input task sequence length increases~\cite{feng2024towards}.

\begin{figure*}[t]
    \centering
    \includegraphics[width=\textwidth]{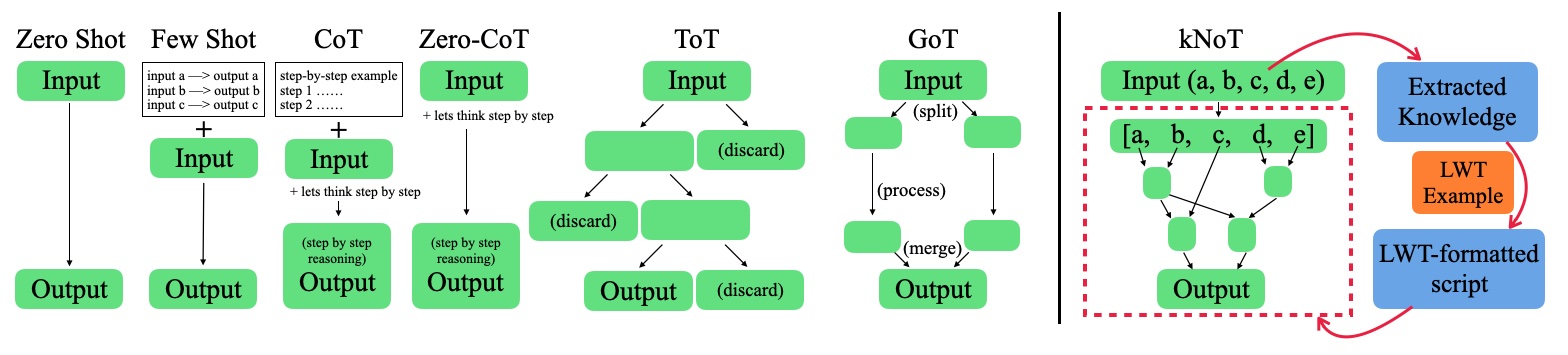}
    \vspace{-6mm}
    \caption{Comparison of \methodfull (\method) to other prompting strategies.}
    \vspace{-3mm}
    \label{fig:compare}
\end{figure*}

Recent studies including Tree of Thoughts (ToT)~\cite{yao2024tree} and Graph of Thoughts (GoT)~\cite{besta2024graph} have improved CoT by modularizing the intermediate thought step into separated LLM inferences to support alternative reasoning structures (see Fig.~\ref{fig:compare}). However, this approach allows each inference to only reference the parent reasoning step as context, potentially generating cascading errors due to the loss of crucial information from earlier ancestor steps~\cite{bao2024llms}. Moreover, these approaches require substantial human effort to design task-specific prompt modules, therefore limiting their generality to support novel task scenarios. Besides, the reasoning structures have limited flexibility, only supporting predefined formats, such as trees for ToT and split-then-merge structure for GoT. Finally, when the processed information increases, the precision will drop because they do not implement elementary operations but process the task as a whole.

To address the aforementioned challenges, this paper aims to propose a \emph{self-guided} approach, enabling a more user-friendly, powerful, and reliable prompt strategy to reduce manual configurations. In essence, our goal is to achieve \textbf{\textit{intelligence amplification}}~\cite{engelbart2023augmenting, carter2017using}, where LLMs not only follow human instructions (e.g., to think step by step) but also actively participate in structuring the content and sequence of instructions to effectively solve a given task. 

Therefore, we present \methodfull (\method), a prompt scheme that \emph{leverages LLM's inherent knowledge to plan solution steps} \textbf{(contribution \#1)} and \emph{autonomously format these steps into a network of thoughts} \textbf{(contribution \#2)}.\footnote{Following~\cite{besta2024graph}, an LLM \textit{thought} denotes an LLM inference performing a reasoning step.} Concretely, given a task instance, \method prompts LLMs to provide a detailed solution plan, which the LLM then executes. This approach not only enhances the LLM's problem-solving capabilities but also significantly reduces the need for task-specific human-engineered prompts. By allowing the LLM to generate its own instruction set, \method achieves greater flexibility across diverse tasks while maintaining high performance. 
 
However, naively asking LLMs to devise a plan directly usually results in vague natural language descriptions that are not executable in their raw form. To address this fundamental issue, we introduce the \textit{\scriptfull (\script)} \textbf{(contribution \#3)}, a structured text format designed to create precise and executable prompt instructions for LLMs. First, \script designs an input field notation to integrate outputs from previous reasoning steps. Visualizing each reasoning step as a vertex, the input field creates an edge between two steps. By enabling connection between any steps, \script supports a generalized \emph{network of thoughts}, transcending the limitations of trees~\cite{yao2024tree} and split-then-merge structures~\cite{besta2024graph}. Second, \script introduces an \textit{indexing} notation to precisely select elements from list-form outputs of reasoning steps, offering fine-grained control over information flow. This flexibility enables \method to construct solution plans consisting of elementary operations, achieving precise and accurate intermediate results \textbf{(contribution \#4)}. This granular approach not only improves the overall accuracy of the problem-solving process but also enables more complex reasoning patterns, allowing LLMs to tackle a wider range of tasks with increased effectiveness and reliability.

To demonstrate the versatility of \method, we experiment on six use cases across three categories, including natural language tasks (Yelp review comprehension and keyword counting), symbolic operation tasks (sorting numbers with duplications and set intersection), and arithmetic tasks (general arithmetic calculation and large digit addition). Our results show that \method significantly outperforms the state of the art across all tasks \textbf{(contribution \#5)}. For example, \method achieves $27\%$ accuracy in sorting $64$ numbers, while all baselines fail entirely with $0\%$ (or $1\%$ for GoT) accuracy. In addition, we also conduct an ablation study to highlight the impact of \method's components \textbf{(contribution \#6)}.

Finally, we introduce a framework for prompt solution procedures (Definition~\ref{def:prompt_adaptation_procedure}) to quantify the human labor required to adapt prompt engineering schemes to new tasks \textbf{(contribution \#7)}. This framework delineates between \textit{constant prompts} that can be reused and \textit{task-specific prompts} that require manual redesign, providing a clear structure for optimizing the prompt engineering process. By applying this framework, we find that \method reduces the need for human labor in engineering \textit{task-specific prompts} by up to $84.4\%$ compared to ToT and up to $87.3\%$ compared to GoT, while maintaining comparable LLM API expenses.

\begin{table}[t]
    \centering
    \resizebox{\linewidth}{!}{
    \begin{tabular}{lcccc}
    \toprule
    \textbf{Prompt Schemes} & \textbf{Mod?} &  \textbf{Net?} & \textbf{Ele?} & \textbf{Amp?} \\ 
    \midrule
    CoT~\cite{wei2022chain} & \faTimes & \faTimes & \faTimes & \faTimes\\
    Zero-CoT~\cite{kojima2022large} & \faTimes & \faTimes & \faTimes & \faTimes\\
    CoT-SC~\cite{wang2022self} & \faTimes & \faTimes & \faTimes & \faTimes\\
    Auto-CoT~\cite{zhang2023automatic} & \faTimes & \faTimes & \faTimes & \faTimes \\
    SoT~\cite{ning2023skeleton} & \faBatteryFull & \faTimes & \faTimes & \faTimes\\
    ToT~\cite{yao2024tree}  & \faBatteryFull & \faBatteryHalf & \faTimes & \faTimes\\
    GoT~\cite{besta2024graph}  & \faBatteryFull & \faBatteryHalf & \faBatteryHalf & \faTimes\\
    \midrule
    \method & \faBatteryFull & \faBatteryFull & \faBatteryFull & \faBatteryFull \\
    \bottomrule
    \end{tabular}
    }
    \caption{Comparison of prompting schemes, with respect to supported reasoning capabilities. \textbf{``Mod?"}: modularized thoughts? \textbf{``Net?"}: network of thoughts? 
    \textbf{``Ele?"}: elementary operations? \textbf{``Amp?"}: intelligence amplification? ``\faBatteryFull": full support, ``\faBatteryHalf": partial support, ``\faTimes": no support.
    }
    \label{tab:qualitative-compare}
\end{table}

\section{Related Work}
Table~\ref{tab:qualitative-compare} compares the different \emph{prompt engineering} methods. 

Chain-of-Thought (CoT) prompting~\cite{wei2022chain} instructs LLMs to solve tasks step-by-step by creating a \textit{chain} of intermediate reasoning \textit{thoughts}. Zero Shot CoT (Zero-CoT)~\cite{kojima2022large} discovers that simply adding ``Let's think step by step'' after the input task improves performance compared to the zero-shot setting. Self-Consistency with CoT (CoT-SC)~\cite{wang2022self} further refines CoT by running it multiple times to generate different answers and selecting the most consistent one as the final result. Auto CoT~\cite{zhang2023automatic} employs Zero-CoT to generate an example for the original CoT operation. Afterward, Skeleton-of-Thoughts (SoT)~\cite{ning2023skeleton} represents an early attempt to \textbf{modularize} each \textit{thought} into separate LLM operations. Recently, Tree of Thoughts (ToT)~\cite{yao2024tree} and Graph of Thoughts (GoT)~\cite{besta2024graph} extend the reasoning structure from a simple chain to more complex tree and split-then-merge frameworks.

However, all the CoT-derived methods face a long context challenges because they require LLMs to reason over an extended stretch of text, including the step-by-step example, the input task, and prior reasoning steps. Thus, the performance deteriorates when the input size increases, as the generated chain of thought can only loosely follow the provided example because the attention mechanism has trouble locating the correct steps in the example~\cite{stechly2024chain}. Recent advancements of SoT, ToT and GoT leverage \textbf{modularization} to alleviate the long context challenge. Nevertheless, these works still exhibit limitations. First, they are confined to predefined structures and do not fully support a \textbf{network of thoughts}. Second, they lack the precision of \textbf{elementary operations}; SoT and ToT processes the entire task, while GoT divides the task into a predefined number of sections. Most importantly, these approaches rely on hand-crafted modules and do not support \textbf{intelligence amplification} with LLMs. For instance, ToT employs a \emph{state evaluator} to decide whether to branch or return to the parent node. GoT uses a \textit{Split Module} to split the task into sections and a \emph{Merge Module} to recombine them. These modules, designed to operate within predefined structures, require significant engineering to adapt to new tasks.

Other research has explored two alternative approaches to enhance LLM task-solving capabilities. One approach uses programming for precise solutions~\cite{chen2023program, gao2023pal}. However, by relying on external interpreters, these methods are restricted to programmable tasks and could not operate on natural language tasks. Another line of research~\cite{yao2022react, shinn2024reflexion, fang2024large, zhang2024proagent} focuses on navigating external environments, iteratively prompting LLMs to determine subsequent exploration steps. Although innovative, these approaches diverge from the focus of fully harnessing LLMs' inherent reasoning capabilities within a self-contained framework and are out of the scope of this work.

\section{Problem Formulation}

    In this work, we address the challenge of devising an effective prompt strategy to harness pretrained LLMs for complex reasoning tasks. To facilitate discussion, we first present the definition of a \textit{prompt engineering problem}.

\begin{definition}[Prompt Engineering Problem]
    \label{def:prompt_engineering_problem}
    Let $\mathcal{L}$ be a pretrained LLM and $\mathcal{T} = \{t_1, \ldots, t_n\}$ be a set of reasoning task categories. For each task $t \in \mathcal{T}$, there is an associated evaluation set $\mathcal{D}_t = \{(\mathbf{q}^t_i, \mathbf{a}^t_i)\}_{i=1}^{N_t}$, where $\mathbf{q}^t_i$ and $\mathbf{a}^t_i$ represent the input query and corresponding answer, respectively, and $N_t$ is the number of samples for task $t$. The Prompt Engineering Problem aims to design a prompt scheme consisting of an algorithmic procedure $\mathcal{A}$, a set of constant prompts $\mathcal{P}_{const}$ applicable across all tasks, and a set of task-specific prompts $\mathcal{P}_{task}^t$ designed for each $t \in \mathcal{T}$, such that
\begin{equation}
    \mathcal{A} (\mathcal{L}, \mathcal{P}_{const}, \mathcal{P}_{task}^t, \mathbf{q}^t_i) = \mathbf{a}^t_i,\,\,\forall t\in\mathcal{T}, i\in [1, N_t].
\end{equation}
\end{definition}
    Note that the LLM $\mathcal{L}$ may be invoked multiple times, taking different combinations of $\mathcal{P}_{const}, \mathcal{P}_{task}^t, \mathbf{q}^t_i$ and potentially the intermediate results as input, according to the algorithmic procedure $\mathcal{A}$. The final output of the prompt scheme must exactly match the correct answer $\mathbf{a}$. Additionally, each task $t \in \mathcal{T}$ may be further divided based on problem size. For instance, in the task of sorting, the problem size corresponds to the number of values to be sorted. This allows for a more granular evaluation of the prompt scheme's effectiveness across varying levels of task difficulty.

    To clarify the cost of solving a prompt engineering problem in terms of manual configurations versus LLM operations, we present the following \textit{prompt solution procedure} framework.

\begin{figure}
    \centering
    \includegraphics[width=.9\linewidth]{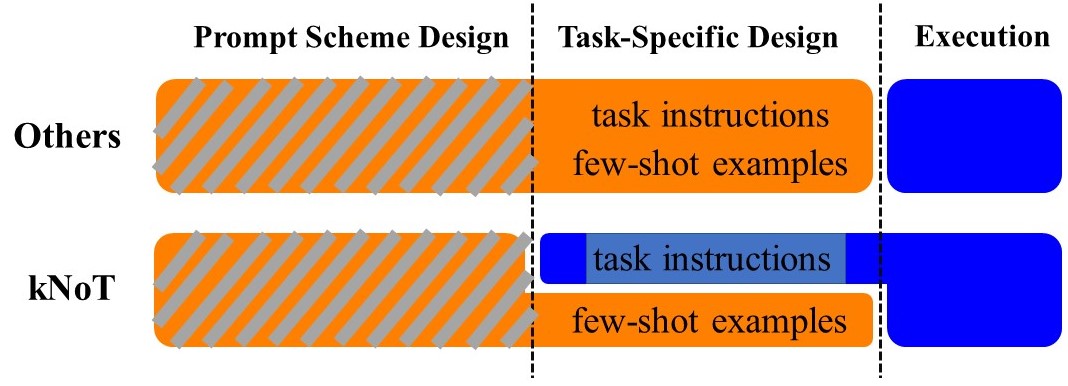}
    \caption{Illustrations of human labor (\textcolor{orange}{orange block} \textcolor{orange}{\faSquare}) and LLM operation (\textcolor{blue}{blue block} \textcolor{blue}{\faSquare}) required for the prompt solution procedure. \textcolor{gray}{Gray stripes} indicate the labor involved in designing \textcolor{gray}{constant prompts}, which do not require redesign.}
    \label{fig:labor-v-llm}
\end{figure}

\begin{figure*}[t]
    \centering
    \includegraphics[width=\textwidth]{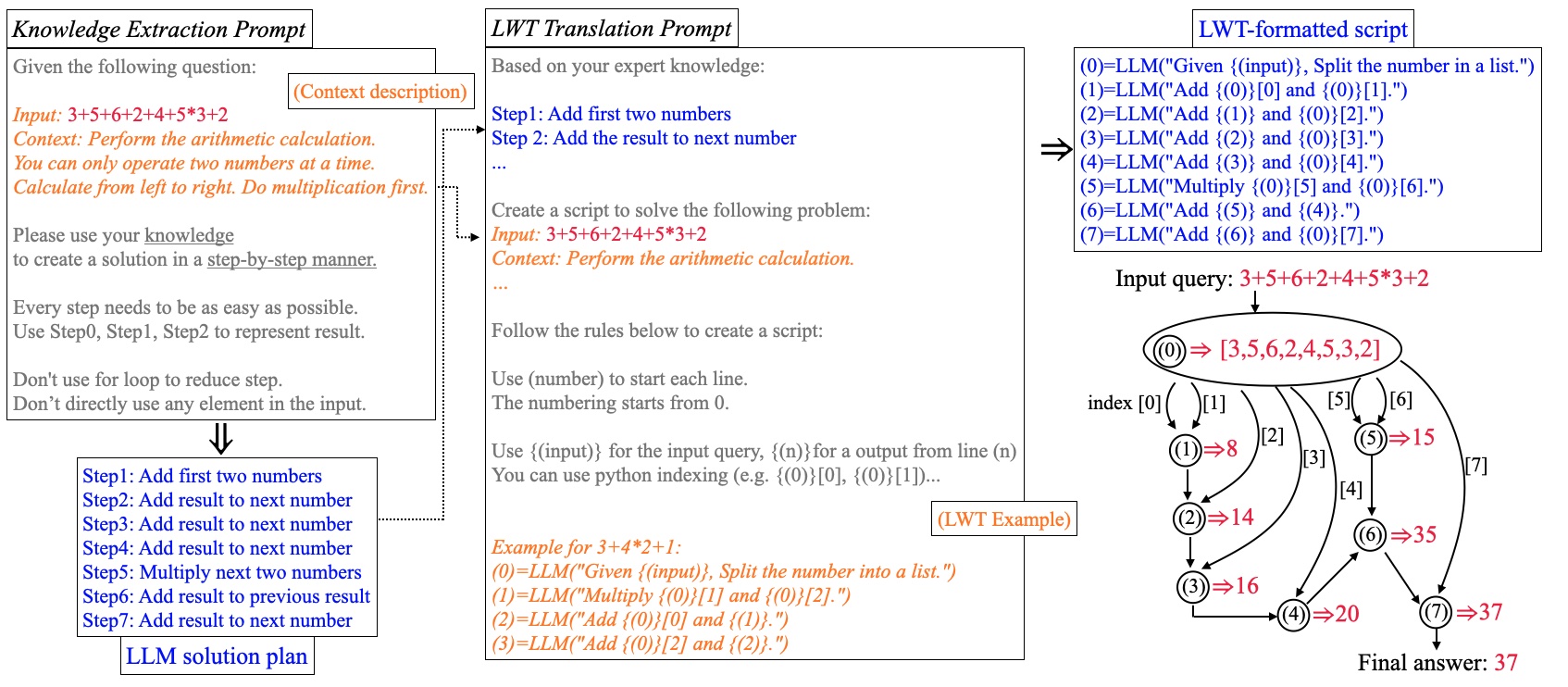}
    \caption{Illustration of the Self-guided \methodfull (\method). The prompts in \method include \textcolor{gray}{\textit{constant prompts}} and \textcolor{orange}{\textit{task-specific prompts}}, comprising a \textcolor{orange}{\textit{context description}} and a \textcolor{orange}{\textit{\script example}}. \method first generates a \textcolor{blue}{LLM solution plan} from the LLM based on the \textcolor{red}{input query}. Then, it transforms the \textcolor{blue}{LLM solution plan} into an \textcolor{blue}{\script-formatted script}. Finally, the bottom-right diagram visualizes the execution of \textcolor{blue}{\script-formatted script}, forming a network of thought processes, where straight or curved arrows \textit{full} or \textit{list-indexed} message passing. Throughout the illustration, $\Rightarrow$ indicates LLM operations.}
    \label{fig:illustration}
\end{figure*}


\begin{definition}[Prompt Solution Procedure]
\label{def:prompt_adaptation_procedure}
    The procedure for solving a prompt engineering problem for a task $t$ consists of the following three stages. 
\begin{enumerate}
    \item \textbf{Prompt scheme design}, which involves designing the algorithmic procedure $\mathcal{A}$ and the constant prompts $\mathcal{P}_{const}$. 
    \item \textbf{Task-specific design} for $\mathcal{P}_{task}^t$, which is further divided into \textbf{task instructions} and \textbf{few-shot examples}.
    \item \textbf{Execution} of the prompt scheme after all prompt engineering efforts are complete to obtain the answers.
\end{enumerate}
\end{definition}

    The second stage, in particular, must be conducted for each novel task and is envisaged as a bottleneck for broader applicability. Particularly, ToT and GoT features multiple modules that all require separate instructions and few-shot examples. In contrast, as shown in Fig.~\ref{fig:labor-v-llm}, \method aims to automatically generate the suitable instructions by LLM's own knowledge, thereby alleviating the required human labor.




\section{\methodfull}
    In this work, we introduce the self-guided \methodfull (\method), a novel prompting scheme designed to address the key challenges of 1) reducing human labor, 2) enabling flexible reasoning structures, and 3) supporting elementary operations. This approach is motivated by the observation that when specifically prompted, LLMs can break down tasks into detailed plans, effectively operating over individual elements of the task input. However, it is difficult for them to utilize their own knowledge when presented with the task (zero-shot, CoT) or the preceding step (ToT, GoT). Similarly, when working on an idea, humans generate new thoughts not only based on the current (ToT and GoT) or all previous thoughts (CoT) but are also guided by a preconceived plan or intuition.

    In particular, \method aims to first leverage LLMs to construct a knowledgeable \textit{solution plan}, then translate the solution plan into an actual executable script. To facilitate the translation from a solution plan to a executable script, \method utilizes the \scriptfull (\script), a specialized text format that enables precise LLM prompt instructions while facilitating sophisticated message passing and indexing designs.

\subsection{\scriptfull (\script)}
    
    \script facilitates flexible reasoning structures by forming a \textit{message-passing network} between different \script-instructions. In particular, it defines ``input fields,'' which receives the output of a previous \script-instruction, forming the (directed) edge of the network. To enable elementary operations, \script further provides indexing capabilities to select individual elements within the output.
    \begin{definition}[\scriptfull (\script)]
    \label{def:lwt}
    An \script-formatted script consists of a list of \script-instructions. Each instruction can reference outputs from previous instructions using the following notation:
    \begin{itemize}
    \item \textsc{\{(n)\}} denotes an input field that receives the entire output of the \textsc{n}$^\text{th}$ \script-instruction.
    \item \textsc{\{(n)\}[m]} selects the \textsc{m}$^\text{th}$ item when \textsc{\{(n)\}} is a list.
    \end{itemize}
    \end{definition}
    Additionally, the notation ``(n)=LLM(...)'' denotes the position of an \script-instruction. Below is an example.
    \mybox{(n)=LLM(``An example \script-instruction with input field \textsc{\{(n)\}} and indexed input field \textsc{\{(n)\}[m]}'')}

\subsection{Detail \method Mechanism}
    Equipped with \script, we present the \textit{\methodfull (\method)}. Fig.~\ref{fig:illustration} illustrates the full \method scheme solving an example arithmetic task. As shown, \method features a three-step process: \textit{1) knowledge extraction}, which extracts LLM's knowledge to generate an \textit{LLM solution plan} $\mathcal{P}$; \textit{2) \script translation}, which converts the plan $\mathcal{P}$ into an executable \textit{\script-formatted script} $\mathcal{S}$; and \textit{3) script execution}, which iteratively executes the \script-instructions in $\mathcal{S}$ to generate the final answer.

    Based on Definition~\ref{def:prompt_adaptation_procedure}, the \script-formatted script $\mathcal{S}$ corresponds to the \textit{task instructions} component of the task-specific prompts $\mathcal{P}^t_{task}$ for a task $t$. Notably, \method automates the generation of $\mathcal{S}$ through its first two steps, significantly reducing the human effort required in prior prompt schemes, as depicted in Fig.~\ref{fig:labor-v-llm}.

    Algorithm~\ref{alg:method} presents a pseudo-code of \method, while Table~\ref{tab:delineation} presents an overview of the prompt instructions and commands used in \method.

\paragraph{Step 1: Knowledge Extraction}
    The first step, illustrated on the left side of Fig.~\ref{fig:illustration}, generates an LLM solution plan using a \kprompt $\mathcal{K}$. This prompt combines \textit{extraction instructions}, the \textit{input query} $\mathbf{q}$, and a \textit{context description} $C$.
    The context description outlines the task's objective and provides solution hints. For instance, in the arithmetic calculation task, the LLM is guided to design a solution using elementary steps, processing two numbers at a time, and to calculate multiplications first.

    Extraction instructions include \textit{elementary commands} and \textit{restriction commands}. Elementary commands direct the LLM to make each step ``as easy as possible'' and to ``Use Step0, Step1, Step2, to represent the results,'' ensuring a step-by-step plan with basic operations. Restriction commands prohibit complex structures like for-loops, as our goal is to obtain a clear, sequential operation order that can be translated into the \script format. Moreover, based on our observations of LLM behavior, we instruct the model to reference input elements positionally rather than verbatim, reducing errors and aligning with \script's indexing notation.
    


\renewcommand{\algorithmiccomment}[1]{{\scriptsize\hfill//\texttt{#1}}}

\begin{algorithm}[t]
    \small
    \caption{\methodfull.}
    \label{alg:method}
    \textbf{Input:} task query $\mathbf{q}$, context $C$, LWT example $\mathcal{E}$, \\
                \> \kprompt $\mathcal{K}$, \tprompt $\mathcal{T}$.
    \textbf{Procedure:}
    \begin{algorithmic}[1]
    \STATE $\mathcal{P}\gets LLM(\mathcal{K}\oplus \{\mathbf{q}^t, C^t\})$ 
    \COMMENT{LLM solution plan}
    \STATE $\mathcal{S}\gets LLM(\mathcal{T}\oplus \{\mathbf{q}^t, C^t, \mathcal{E}, \mathcal{P}\})$ \COMMENT{\script-formatted script}
    \STATE Initiate $L$ as empty list \COMMENT{Script execution}
    \FOR{\script-instruction \textbf{in} \script-formatted script $\mathcal{S}$}
        \FOR{input\_field $n$ \textbf{in} \script-instruction}
            \IF{input\_field \textbf{is not} indexed}
                \STATE \script-instruction $\gets$ \script-instruction$\oplus L[n]$
            \ELSIF{input\_field \textbf{is} indexed by $m$}
                \STATE \script-instruction $\gets$ \script-instruction$\oplus L[n][m]$
            \ENDIF
        \ENDFOR
        \STATE output $\gets$ $LLM(\text{\script-instruction})$
        \STATE $L$.append(output)
    \ENDFOR
    \RETURN output
    \end{algorithmic}
\end{algorithm}

\begin{table}[t]
    \centering
    \resizebox{\linewidth}{!}{
    \begin{tabular}{l|cc}
    \toprule
        & Constant Prompt $\mathcal{P}_{const}$ & Task-specific Prompt $\mathcal{P}_{task}^t$\\
        \midrule
        Step 1 & Extraction instructions  & \textcolor{orange}{Context description $C$}\\
        Step 2 &  Translation instructions & \textcolor{orange}{\script Example $\mathcal{E}$}\\
        Step 3 & -- & \textcolor{blue}{\script-formatted script $\mathcal{S}$} \\
        \bottomrule
    \end{tabular}
    }
    \caption{Delineation of constant versus task-specific prompts for \method with task $t$. We highlight manual labor in \textcolor{orange}{orange} and LLM automated results in \textcolor{blue}{blue}.}
    \vspace{-4mm}
    \label{tab:delineation}
\end{table}

\paragraph{Step 2: \script Translation}

    As depicted in the middle of Fig.~\ref{fig:illustration}, step 2 focuses on translating the solution plan $\mathcal{P}$ into a \script-formatted script $\mathcal{S}$ using a \tprompt $\mathcal{T}$. This prompt integrates \textit{translation instructions}, the solution plan $\mathcal{P}$, an \script example $\mathcal{E}$, along with $\mathbf{q}$ and $C$ for proper context. The translation instructions direct the LLM to number the \script-instructions and provide an explanation of the \script format. The \script example presents a solution to a downsized version of the same task category. For instance, when calculating arithmetic sequences of $16$, $32$, or $64$ numbers, we provide an \script example for a sequence of $4$ numbers. A design approach we adopt in most \script examples is to make the first operation split the input sequence into a list. This approach facilitates elementary operations, allowing subsequent \script-instructions to access individual elements by indexing the correct position from the output of the first instruction.

\paragraph{Step 3: Script Execution}

    Finally, the \script-formatted script $\mathcal{S}$ is executed sequentially. As shown in Algorithm~\ref{alg:method} (Lines 3-11), an empty list $L$ is initialized to store the outputs. For each \script-instruction in $\mathcal{S}$, \method formats its input fields. Non-indexed fields append the entire output from $L$, denoted as $L[n]$, where $n$ is the index of the previous output. If an input field is indexed by $m$, it appends the $m$th element, denoted as $L[n][m]$. The formatted \script-instruction is then executed by an LLM, and the output is appended to $L$. This process allows subsequent instructions to access specific parts or entire previous outputs. The final instruction's output is returned as the answer to query $\mathbf{q}$.
\subsection{Example Use Cases}
    We demonstrate six use cases of \method across three categories. In each case, we provide basic \textit{hints} through the context description $C$ and model a solution strategy with the \script example $\mathcal{E}$, showcasing the flexibility of \script.\footnote{For instance, the context description for sorting simply states \textit{``Sort in ascending order. You can use counting sort.''} Due to space constraints, we provide the exact prompt designs in Appendix~\ref{appsub:knot_task_specific_prompts}.}
    
    First, for \textit{natural language tasks}, we address \textbf{1) Yelp review comprehension}, which involves counting the number of positive reviews within a batch of mixed reviews from the Yelp dataset~\cite{yelp_dataset}, and \textbf{2) keyword counting}, where the goal is to identify all keywords in an article that belong to the same category (e.g., country names). In both tasks, \method splits the individual reviews (or sentences) from the batch (or article) and analyzes each one sequentially, aggregating the results as they proceed. By modularizing each analysis, \method ensures precise and accurate outcomes.

    Second, for \textit{symbolic operation tasks}, we solve \textbf{3) sorting} numbers with duplicates and \textbf{4) set intersection} between two sets. For sorting, \method employs counting sort, as it demands the least context size to maintain intermediate results. The \script example for sorting models a strategy that first initializes an array of size $10$, and then sequentially updates the count of each number. For set intersection, \method checks whether each element in the first set exists in the second set, one by one. Specifically, the \script-instruction outputs the element if it exists in the second set, and nothing otherwise. A final \script-instruction aggregates the results from all preceding steps into a single array.
            
    Lastly, we tackle arithmetic tasks, including \textbf{5) arithmetic calculations} involving addition, multiplication, subtraction, and division for a sequence of two-digit numbers, and \textbf{6) large-digit addition} between two multi-digit numbers. For the former, \method first splits the numbers into an array, discarding the operators which would be carried out in the corresponding \script-instruction steps within the \script-formatted script $\mathcal{S}$. For the latter, \method splits the two numbers into individual digits and begins the addition from the least significant digit. It retains the units digit of the result and performs a division by $10$ to determine whether to carry over a digit, which is passed to the next step. This process repeats until \method reaches the most significant digit. Finally, a \script-instruction concatenates all the units digits from each step to obtain the final answer.


\begin{table*}[t]
    \centering
    \resizebox{\textwidth}{!}{
    \begin{tabular}{lcccccccccccccc}
    \toprule
     & \multirow{3}{*}{
     \textbf{Yelp}
     }   & \multirow{3}{*}{\shortstack{\textbf{keyword}\\\textbf{counting}}} & \multicolumn{3}{c}{\textbf{sorting}} & \multicolumn{3}{c}{\textbf{set intersection}} & \multicolumn{3}{c}{\textbf{arithmetic}} & \multicolumn{3}{c}{\textbf{large-digit}} \\
     \cmidrule(lr){4-6} \cmidrule(lr){7-9}\cmidrule(lr){10-12}\cmidrule(lr){13-15}
     & & & 16 & 32 & 64
     & 32 & 64 & 128 & 8 & 16 & 32 & 8 & 16 & 32\\
    \midrule
    Zero Shot & $28\%$ & $0\%$ & $86\%$ & $0\%$ & $0\%$ & $0\%$ & $0\%$ & $0\%$ & $16\%$ & $0\%$ & $0\%$ & $40\%$ & $20\%$ & $12\%$\\
    Few Shot & $32\%$ & $0\%$ & $88\%$ & $1\%$ & $0\%$ & $2\%$ & $0\%$ & $0\%$ & $22\%$ & $0\%$ & $0\%$ & $43\%$ & $25\%$ & $20\%$\\
    Zero-CoT~\cite{kojima2022large} & $40\%$ & $0\%$ & $92\%$ & $0\%$ & $0\%$ & $2\%$ & $0\%$ & $0\%$ & $26\%$ & $0\%$ & $0\%$ & $45\%$ & $25\%$ & $21\%$\\
    CoT~\cite{wei2022chain} & $46\%$ & $0\%$ & $94\%$ & $0\%$ & $0\%$ & $5\%$ & $0\%$ & $0\%$ & $36\%$ & $14\%$ & $0\%$ & $50\%$ & $27\%$ & $23\%$ \\
    CoT-SC~\cite{wang2022self} & $48\%$ & $0\%$ & $98\%$ & $0\%$ & $0\%$ & $7\%$ & $0\%$ & $0\%$ & $36\%$ & $12\%$ & $0\%$ & $52\%$ & $28\%$ & $24\%$\\
    ToT~\cite{yao2024tree} & $40\%$ & $1\%$ & $100\%$ & $12\%$ & $0\%$ & $29\%$ & $0\%$ & $0\%$ & $42\%$ & $8\%$ & $0\%$ & $40\%$ & $20\%$ & $5\%$\\
    GoT~\cite{besta2024graph} & $52\%$ & $25\%$ & $100\%$ & $31\%$ & $1\%$ & $44\%$ & $7\%$ & $1\%$ & $12\%$ & $5\%$ & $0\%$ & $9\%$ & $0\%$ & $0\%$\\
    \midrule
    \textbf{\method} & $\mathbf{75\%}$ & $\mathbf{45\%}$ & $\mathbf{100\%}$ & $\mathbf{92\%}$ & $\mathbf{27\%}$ & $\mathbf{93\%}$ & $\mathbf{32\%}$ & $\mathbf{20\%}$ & $\mathbf{90\%}$ & $\mathbf{32\%}$ & $\mathbf{10\%}$ & $\mathbf{98\%}$ & $\mathbf{88\%}$ & $\mathbf{56\%}$\\
    \bottomrule
    \end{tabular}
    }
    \caption{Comparison of prompt scheme accuracy over different use cases.}
    \label{tab:full}
\end{table*}


\subsection{Qualitative Analysis of \method's Advantages}

    We focus our comparison with the state-of-the-art methods GoT~\cite{besta2024graph} and ToT~\cite{yao2024tree}. The advantages of \method are three-fold. First, \method achieves \textbf{intelligence amplification} by autonomously generating the \textit{\script-formatted script} $\mathcal{S}$ for an input task $\mathbf{q}$. \method only requires a single set of \script example $\mathcal{E}$ for each task type, significantly reducing the manual configurations required. In contrast, both ToT and GoT rely on multiple hand-crafted modules. E.g., ToT uses the (process) module while GoT uses (split), (process), and (merge) modules. Each module's distinct functionality requires customization to the specific task, necessitating separate \textit{task-specific prompts}, including instructions and few-shot examples, incurring significant human configurations. 

    Second, \method constructs a more intricate \textbf{network of thoughts}. Unlike previous approaches that follow preset structures such as a single chain (CoT), a tree (ToT), or a split-then-merge structure (GoT), \method leverages \script's built-in flexibility to create sophisticated networks that coordinate various reasoning steps. Moreover, by exploiting LLM to design the network, \method effectively customizes the reasoning structure for each task instance, generating unique \script-formatted scripts (see Appendix~\ref{appsub:customized}).

Lastly, \method utilizes \textbf{elementary operations} to achieve precise and accurate intermediate results. Enabled by \script's indexing capability, \method breaks down the input task query into smaller elements for fine-grained processing, instead of operating on the entire task sequence (such as in CoT, ToT) or sections of the task sequence (such as in GoT). 

\begin{table*}[t]
  \begin{minipage}{0.48\textwidth}
    \centering
    \resizebox{\linewidth}{!}{
    \begin{tabular}{lcccccccccccccccc}
    \toprule
     & \1 &  \2 & \3  & \1 + \2 & \1 + \3 & \2 + \3 & \method \\
    \midrule
    Yelp & $38\%$ & $91\%$ & $74\%$ & $96\%$ & $90\%$ & $96\%$ & $100\%$\\
    keyword & $59\%$ & $94\%$ & $67\%$ & $94\%$ & $88\%$ & $98\%$ & $100\%$\\
    sorting & $76\%$ & $90\%$ & $84\%$ & $92\%$ & $90\%$ & $96\%$ & $100\%$\\
    set operation & $68\%$ & $91\%$ & $75\%$ & $93\%$ & $92\%$ & $97\%$ & $100\%$\\
    arithmetic & $88\%$ & $92\%$ & $94\%$ & $95\%$ & $99\%$ & $98\%$ & $100\%$\\
    large digit & $51\%$ & $91\%$ & $60\%$ & $94\%$ & $85\%$ & $98\%$ & $100\%$\\
    \bottomrule
    \end{tabular}
    }
    \vspace{-2mm}
    \caption{Ablation results of \kprompt.}
    \vspace{-2mm}
    \label{tab:ablation1}
  \end{minipage}
  \hfill
  \begin{minipage}{0.48\textwidth}
    \centering
    \resizebox{\linewidth}{!}{
    \begin{tabular}{lcccccccccccccccc}
    \toprule
     & \4 &  \5 & \6  & \4 + \5 & \4 + \6 & \5 + \6 & \method \\
    \midrule
    Yelp & $0\%$ & $0\%$ & $90\%$ & $0\%$ & $94\%$ & $93\%$ & $100\%$\\
    keyword & $0\%$ & $0\%$ & $92\%$ & $0\%$ & $96\%$ & $97\%$ & $100\%$\\
    sorting & $0\%$ & $0\%$ & $88\%$ & $0\%$ & $95\%$ & $92\%$ & $100\%$\\
    set operation & $0\%$ & $0\%$ & $91\%$ & $0\%$ & $97\%$ & $99\%$ & $100\%$\\
    arithmetic & $0\%$ & $0\%$ & $94\%$ & $0\%$ & $96\%$ & $98\%$ & $100\%$\\
    large digit & $0\%$ & $0\%$ & $92\%$ & $0\%$ & $93\%$ & $97\%$ & $100\%$\\
    \bottomrule
    \end{tabular}
    }
    \vspace{-2mm}
    \caption{Ablation results of \tprompt.}
    \vspace{-2mm}
    \label{tab:ablation2}
  \end{minipage}
\end{table*}



\section{Experiment}
\label{sec:experiment}

\subsection{Setup}
    We outline the setup for task queries $\mathbf{q}$ for each use case (actual samples are provided in Appendix~\ref{appsub:task_query_example}). \textbf{1) Yelp} presents a randomly mixed batch of 5-star (positive) and 1-star (negative) reviews from the Yelp dataset~\cite{yelp_dataset}. \textbf{2) Keyword} provides an article with $14-20$ sentences. \textbf{3) Sorting} presents an array single-digit numbers with duplicates. \textbf{4) Set intersection} uses two lists of double-digit numbers. \textbf{5) Arithmetic} presents a sequence of double-digit numbers involving addition, multiplication, subtraction, and division.\footnote{Floating point values are rounded to two decimal places.} \textbf{6) Large-digit} presents a two-number addition problem. For symbolic operations and arithmetic tasks, we designate three different problem sizes to test the scalability of prompt schemes. For each problem size, we prepare 100 sample queries for evaluation.
    
    We compare \method with eight prompt schemes, including the Graph of Thoughts (GoT)~\cite{besta2024graph}, Tree of Thoughts (ToT)~\cite{yao2024tree}, Self-consistent Chain of Thoughts (CoT-SC)~\cite{wang2022self}, Chain-of-thought (CoT)~\cite{wei2022chain}, Zero-shot CoT (Zero CoT)~\cite{kojima2022large} as well as the basic Few Shot and Zero Shot prompting. We follow the prompt scheme design from the original source codes and manually prepare task-specific designs for task that were not covered by the source codes (see Appendix~\ref{appsub:manual_prompt_baseline}). For \method, we directly leverage the \script-formatted script generated with GPT-4o as the task instructions and do not conduct any additional manual adjustments. We set the temperature to $0.0$ and use GPT-3.5-turbo $16$k context window size and $4$k maximum output tokens for all prompt scheme execution unless stated otherwise. 
    
\begin{figure}[t]
    \centering
    \includegraphics[width=.8\linewidth]{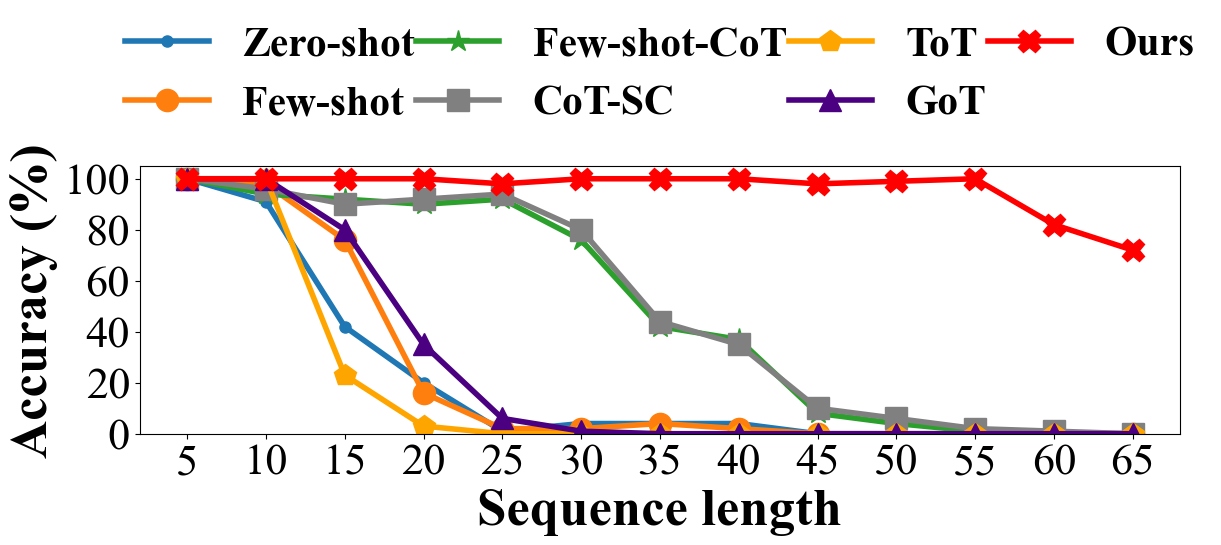}
    \caption{Scalability analysis across varying lengths of addition sequences.}
    \vspace{-6mm}
    \label{fig:add}
\end{figure}
\begin{table*}[t]
  \begin{minipage}{0.48\textwidth}
    \centering
    \resizebox{\linewidth}{!}{
    \begin{tabular}{lcccccc}
    \toprule
      & \textbf{Yelp} & \textbf{keyword} & \textbf{sorting} & \textbf{set} & \textbf{arithmetic} & \textbf{large-digit}  \\
    \midrule
    ToT & \$0.760 & \$1.041 & \$0.513 & \$0.320 & \$0.251 & \$0.292\\
    Accuracy & $100\%$ & $100\%$ & $100\%$ & $99\%$ & $100\%$ & $100\%$\\
    \midrule
    GoT & \$0.371 & \$0.647 & \$0.314 & \$0.132 & \$0.112 & \$0.156\\
    Accuracy & $100\%$ & $99\%$ & $100\%$ & $100\%$ & $100\%$ & $100\%$\\
    \midrule
    \textbf{\method} & \textbf{\$0.245} & \textbf{\$0.526} & \textbf{\$0.267} & \textbf{\$0.115} & \textbf{\$0.091} & \textbf{\$0.085} \\
    Accuracy & $100\%$ & $100\%$ & $100\%$ & $100\%$ & $100\%$ & $100\%$\\
    \bottomrule
    \end{tabular}
    }
    \caption{LLM API Cost for high performance  with GPT-4o.}
    \vspace{-4mm}
    \label{tab:cost}
  \end{minipage}
  \hfill
  \begin{minipage}{0.48\textwidth}
    \centering
    \resizebox{\linewidth}{!}{
    \begin{tabular}{lcccccc}
    \toprule
      & \textbf{Yelp} & \textbf{keyword} & \textbf{sorting} & \textbf{set} & \textbf{arithmetic} & \textbf{large-digit}  \\
    \midrule
    ToT & 3845 & 4986 & 3240 & 4114 & 1609 & 2143\\
    GoT & 4504 & 7534 & 4640 & 5059 & 2434 & 3568\\
    \midrule
    \textbf{\method} & \textbf{841} & \textbf{1033} & \textbf{1049} & \textbf{641} & \textbf{1542} & \textbf{945}\\
    (v. ToT) & $-78.1\%$ & $-79.2\%$ & $-67.6\%$ & $-84.4\%$ & $-4.2\%$ & $-55.9\%$\\
    (v. GoT) & $-81.3\%$ & $-86.3\%$ & $-77.3\%$ & $-87.3\%$ & $-36.6\%$ & $-73.5\%$\\
    \bottomrule
    \end{tabular}
    }
    \caption{Task-specific prompt character counts.}
    \vspace{-4mm}
    \label{tab:human_character}
  \end{minipage}
\end{table*}



\subsection{Results Analysis}
\label{subsec:main-results}

Table~\ref{tab:full} presents a comprehensive summary of the results across all tasks. As shown, \method consistently outperforms all baselines by a significant margin across diverse reasoning tasks. This advantage is especially pronounced in larger problem sizes, such as arithmetic calculations with $64$ numbers or set intersection with $64$ and $128$ numbers, where most baselines drop to zero (with GoT achieving only single-digit accuracy), while \method maintains a strong double-digit performance.

In natural language tasks, \method obtains superior performance because it adopts a strategy to split articles into sentences (or batch of reviews into individual reviews) to precisely analyze each part. Although GoT also employs sentence splitting, its pre-determined merge module operation proves less precise than \method's approach to aggregate each sentence one by one. In contrast, CoT and ToT process entire articles at once, leading to failure due to the LLM's inability to handle such long sequences effectively.

For symbolic operation tasks, the flexibility of \script allows \method to operate with the most concise approach in terms of LLM context length. In particular, for the sorting task, \method utilizes bucket sort to address the problem, as other sorting methods, like merge sort (used by GoT), tend to increase the length of the input with each sorting step. As a result, the sorted array becomes progressively longer until it exceeds the context length that the LLM can process correctly. This is why other baselines fail when the array size increases to 16 and 32. In the set operation task, \method checks whether each number in one set is present in the other set. This precise and straightforward approach in each step allows \method to outperform all baselines across all length settings. Other baselines, like CoT and ToT, attempt to find the intersection of two sets in a single run, making it difficult for the LLM to accurately identify all intersecting numbers. GoT tries to improve accuracy by splitting the sets into smaller subsets, but this approach lacks precision, resulting in lower accuracy compared to \method.

Although GoT generally outperforms other baselines, it falters in arithmetic tasks due to its split-then-merge structure, which is antithetical to the reasoning required for these problems. Arithmetic tasks demand separate consideration of each number based on its relation to nearby numbers, whereas GoT arbitrarily splits the task sequence evenly and relies on a single manually prepared merge structure. In contrast, \method automatically designs customized \script-instructions for each intermediate calculation, enabling more effective problem-solving. Besides, Fig.~\ref{fig:add} provides a scalability test over different lengths of addition sequences, proving the generalizability of \method.


\subsection{Ablation Study}
Tables~\ref{tab:ablation1} and~\ref{tab:ablation2} present ablation results. Specifically, \1, \2, and \3 in Table~\ref{tab:ablation1} denote the context description $C$, the elementary commands, and the restriction commands within the \kprompt. Similarly, \4, \5, and \6 in Table~\ref{tab:ablation2} denote the context description $C$, the translation instructions, and the \script example $\mathcal{E}$ within the \tprompt. To clearly demonstrate the impact of ablating specific components, we use GPT-4o for script execution, enabling \method to achieve 100\% accuracy across all tasks.

From Table~\ref{tab:ablation1}, we observe that the elementary commands alone yield decent accuracy of over $90\%$. However, all ablation versions pale in comparison with the full \method, illustrating the importance of all three components. Table~\ref{tab:ablation2} shows more drastic differences, with the \script example $\mathcal{E}$ emerging as a critical component. Nonetheless, the other two components still contribute positively, pushing the performance to $100\%$ under GPT-4o.


\subsection{Cost Analysis}
We focus our comparison with the state-of-the-art methods ToT and GoT. First, Table~\ref{tab:cost} presents the LLM API costs for high-performance executions using GPT-4o, where all methods achieve near $100\%$ accuracy. In this scenario, \method consistently incurs the lowest cost per task query in USD, as it leverages elementary operations that require the fewest resources. In contrast, ToT incurs the highest cost by continuously operating over the entire task, while GoT ranks second, as it splits each task into equally sized chunks.

Furthermore, Table~\ref{tab:human_character} compares the number of characters required in the task-specific prompts for each task.\footnote{Table~\ref{tab:human} provides comparison with tokens in Appendix~\ref{appsub:more-experiment}.} On average, \method requires only $600$ tokens per task---a $59\%$ reduction compared to ToT's $1467$ tokens and a $68\%$ reduction compared to GoT's $1893$ tokens.




\section{Conclusion}

In this work, we introduced \method, a novel prompt scheme designed to achieve intelligence amplification, leveraging LLMs to draft and prepare precise \script-formatted scripts. \method demonstrates superior performance across a variety of reasoning tasks while reducing both the human labor involved in the task-specific prompt redesigns and the associated LLM operation costs. Future work aims to extend \method's framework to support a wider range of reasoning task categories~\cite{sun2023survey}, and to design more sophisticated prompt schemes that incorporate self-guided revisions for the \script-formatted scripts.

\bibliography{main.bib}

\clearpage
\appendix

\section{Additional Experiment Results}
\label{app:additional}

\subsection{Customized \script-formatted scripts}
\label{appsub:customized}

\method tailors the \script-formatted scripts automatically according to the input task query $\mathbf{q}$. We provide two examples under the arithmetic task. Below is an query instance, followed by the \script-formatted script generated by \method for this instance.

\mybox{5*5/5*4+8-8+3*9}

\mybox{
(0)=LLM("Given \{(input)\}, Split the numbers without operators. Only output list.") \\
(1)=LLM("Multiply(\{(0)\}[0], \{(0)\}[1]). Only output number. If contains floating point, round to two decimal places.") \\
(2)=LLM("Divide(\{(1)\}, \{(0)\}[2]). Only output number. If contains floating point, round to two decimal places.") \\
(3)=LLM("Multiply(\{(2)\}, \{(0)\}[3]). Only output number. If contains floating point, round to two decimal places.") \\
(4)=LLM("Add(\{(3)\}, \{(0)\}[4]). Only output number. If contains floating point, round to two decimal places.") \\
(5)=LLM("Minus(\{(4)\}, \{(0)\}[5]). Only output number. If contains floating point, round to two decimal places.") \\
(6)=LLM("Multiply(\{(0)\}[6], \{(0)\}[7]). Only output number. If contains floating point, round to two decimal places.") \\
(7)=LLM("Add(\{(5)\}, \{(6)\}). Only output number. If contains floating point, round to two decimal places.") \\
}

Below is another query instance, followed by the \script-formatted script generated by \method for this instance.

\mybox{1+5+7+8+2-8-7*7}

\mybox{
(0)=LLM("Given \{(input)\}, Split the numbers without operators. Only output list.") \\
(1)=LLM("Multiply(\{(0)\}[6], \{(0)\}[7]). Only output number. If contains floating point, round to two decimal places.") \\
(2)=LLM("Add(\{(0)\}[0], \{(0)\}[1]). Only output number. If contains floating point, round to two decimal places.") \\
(3)=LLM("Add(\{(2)\}, \{(0)\}[2]). Only output number. If contains floating point, round to two decimal places.") \\
(4)=LLM("Add(\{(3)\}, \{(0)\}[3]). Only output number. If contains floating point, round to two decimal places.") \\
(5)=LLM("Add(\{(4)\}, \{(0)\}[4]). Only output number. If contains floating point, round to two decimal places.") \\
(6)=LLM("Minus(\{(5)\}, \{(0)\}[5]). Only output number. If contains floating point, round to two decimal places.") \\
(7)=LLM("Minus(\{(6)\}, \{(1)\}). Only output number. If contains floating point, round to two decimal places.") \\
}

It is worth noting that the task-specific prompts of \method, including context description $C$ and \script Example $\mathcal{E}$ are the same in both cases. As exemplified by the two cases above, \method achieves \textit{intelligence amplification} by autonomously customizing the solution plan for the specific task instance.

\subsection{Additional Experiment Results}
\label{appsub:more-experiment}
\paragraph{Extra Arithmetic Calculation Experiments}
In the primary arithmetic calculation task, we tested all four arithmetic operations: addition, multiplication, subtraction, and division. Here, we present additional experimental results focused on ``pure addition" and calculations involving only ``addition and multiplication." As shown in Table~\ref{tab:additional_arithmetic}, \method consistently outperforms all the other baselines in these scenarios as well.

\begin{table}[t]
    \centering
    \resizebox{\linewidth}{!}{
    \begin{tabular}{lcccccc}
    \toprule
     \multirow{3}{*}{method} & \multicolumn{3}{c}{\textbf{addition}} & \multicolumn{3}{c}{\textbf{addition and multiplication}} \\
     \cmidrule(lr){2-4} \cmidrule(lr){5-7}
     & 8 & 16 & 32 & 8 & 16 & 32 \\
    \midrule
    Few Shot & 0.94 & 0.76 & 0.04 & 0.24 & 0 & 0\\
    CoT & 0.98 & 0.92 & 0.64 & 0.68 & 0.35 & 0.02\\
    CoT-SC & 1 & 0.93 & 0.62 & 0.69 & 0.33 & 0.04\\
    ToT & 1 & 0.23 & 0 & 0.52 & 0.19 & 0.06\\
    GoT & 1 & 0.8 & 0.01 & 0.28 & 0.01 & 0\\
    \midrule
    \textbf{\method} & \textbf{1} & \textbf{1} & \textbf{1} & \textbf{0.98} & \textbf{0.56} & \textbf{0.2}\\
    \bottomrule
    \end{tabular}
    }
    \caption{More experiment results of arithmetic calculation with pure addition as well as ``addition and multiplication.''}
    \label{tab:additional_arithmetic}
\end{table}

\paragraph{Task-specific prompt token counts} We calculate the token counts in Table~\ref{tab:human} in order to facilitate API cost considerations.

\begin{table}[t]
    \begin{minipage}{0.48\textwidth}
    \centering
    \resizebox{\linewidth}{!}{
    \begin{tabular}{lcccccc}
    \toprule
      & \textbf{Yelp} & \textbf{keyword} & \textbf{sorting} & \textbf{set} & \textbf{arithmetic} & \textbf{large digit}  \\
    \midrule
    ToT & 1843 & 1274 & 2014 & 2064 & 550 & 596\\
    GoT & 2032 & 1803 & 2534 & 2464 & 791 & 853\\
    \midrule
    \textbf{\method} & \textbf{254} & \textbf{278} & \textbf{352} & \textbf{209} & \textbf{395} & \textbf{356}\\
    \bottomrule
    \end{tabular}
    }
    \caption{Task-specific prompt token counts.}
    \label{tab:human}
  \end{minipage}
\end{table}

\section{\method Implementation Details}
\label{app:knot_details}
In the following, we provide the detail prompt utilized in \method.

\subsection{Constant Prompts of \method}

\paragraph{\kprompt} The following is the knowledge extraction prompt:

\mybox{
Given the following question:
\textcolor{blue}{context description}\\
The Input section is the input query.\\ 
The Context section is the goal we want to achieve. \\
Please use your knowledge to create a solution by step-by-step manner without any numbers. \\
Every step need to be as easy as possible. \\
Use Step0, Step1, Step2 to represent result.\\
Don't use for loop to reduce step. \\
Don't directly use any element in the input. 
}

\paragraph{\tprompt} The following is the \script translation prompt:

\mybox{
Based on your expert knowledge \n{generated knowledge} and the above example, create a script to solve the following question:\\
Input: \textcolor{red}{...}\\
Context: \textcolor{orange}{...}\\
\n{context description}
The Input section is the input query. The Context section is the goal we want to achieve.
You have to follow the rules to create a script. \\
This script should be numbered and contains several instruction to be called line-by-line in a sequential order. \\
Use (number) to represent each line. \\
The line numbering starts from 0. \\
You can use LLM Inference: use LLM("Your Instruction") to find the answer. \\
Use \{(index)\} to represent the variable you want to replace with previous result. \\
Use \{(input)\}, \{(Set1)\}, ... to represent input, not allow to directly use numbers. \\
Use python indexing to get the element in the list (E.g. \{(0)\}[0], \{(0)\}[1]). \\
Do not directly use numbers. \\
Here is one example. \\
\textcolor{blue}{\script-script example}
}

\subsection{Task Specific Prompts of \method}
\label{appsub:knot_task_specific_prompts}

\paragraph{Yelp Review Comprehension} 
Every input query in Yelp review comprehension contains 10 reviews. \method considers each review one by one, determining whether it is positive or negative. Then it combines all previous results by counting the number of positive reviews in the final step. The context description and the \script-script example are as the following.

    \mybox{Context: Output how many positive reviews in the input. Check every review one by one in the input.}

    \mybox{(0)=LLM("Check the following review is Positive or Negative: \{(input)\}[0].") \\
    (1)=LLM("Check the following review is Positive or Negative: \{(input)\}[1].") \\
    (2)=LLM("Check the following review is Positive or Negative: \{(input)\}[2].") \\
    ... \\
    (length-1)=LLM("Check the following review is Positive or Negative: \{(input)\}[length-1].") \\
    (length)=LLM("[\{(0)\}, \{(1)\}, \{(2)\}, \{(3)\}, \{(4)\}, \{(5)\},.... ,\{(length-1)\}], output the number of Positive.")}

\paragraph{Keyword Counting} Every input in keyword counting contain an article with 14 to 20 sentences. \method divide the whole article into sentences and check the keyword one by one. In the final step, we will output an array combined with previous results. The context description and the \script-script example are as the following.

\mybox{Context: Output all words about countries in the article. You can separate article into sentences first.}

\mybox{(0)=LLM("Split the following article into sentences: '\{(input)\}'. Output an array.") \\
(1)=LLM("Extract all country names (no continents) in the order of their appearance from the following sentence (repeated is allowed): "\{(0)\}[0]"  Output [] if not exist any country.") \\
(2)=LLM("Extract all country names (no continents) in the order of their appearance from the following sentence (repeated is allowed): "\{(0)\}[1]"  Output [] if not exist any country.") \\
(3)=LLM("Extract all country names (no continents) in the order of their appearance from the following sentence (repeated is allowed): "\{(0)\}[2]"  Output [] if not exist any country.") \\
... \\
(20)=LLM("Extract all country names (no continents) in the order of their appearance from the following sentence (repeated is allowed): "\{(0)\}[19]"  Output [] if not exist any country.") \\
(21)=LLM("Combine \{(1)\}, \{(2)\}, \{(3)\}, \{(4)\}, \{(5)\}, \{(6)\}, \{(7)\}, \{(8)\}, \{(9)\}, \{(10)\}, \{(11)\}, \{(12)\}, \{(13)\}, \{(14)\}, \{(15)\}, \{(16)\}, \{(17)\}, \{(18)\}, \{(19)\}, \{(20)\} in one array. Repeated is allowed.")}

\paragraph{Sorting} \method uses counting sort. The context description and the \script-script example are as the following.

\mybox{Context: Sort input in ascending order. You can use counting sort.}

\mybox{(0)=LLM("Initialize an array of size 10 to zero.") \\
(1)=LLM("Increment the count at index \{(input)\}[0] in \{(0)\} (index start from 0). Only output updated array.") \\
(2)=LLM("Increment the count at index \{(input)\}[1] in \{(1)\} (index start from 0). Only output updated array.") \\
... \\
(16)=LLM("Increment the count at index \{(input)\}[15] (start from 0) in \{(15)\}. Only output updated array.") \\
... \\
(length+1)=LLM("Convert \{(length)\} in English. Output an array.") \\
(length+2)=LLM("The array should contain \{(length+1)\}[0] 0s, \{(length+1)\}[1] 1s, \{(length+1)\}[2] 2s, \{(length+1)\}[3] 3s, \{(length+1)\}[4] 4s. Output in array format.") \\
(length+3)=LLM("The array should contain \{(length+1)\}[5] 5s, \{(length+1)\}[6] 6s, \{(length+1)\}[7] 7s, \{(length+1)\}[8] 8s, \{(length+1)\}[9] 9s. Output in array format.") \\
(length+4)=LLM("Combine \{(length+2)\} and \{(length+3)\} in ascending order. Only output array.")}

\paragraph{Set Operation} The context description and the \script-script example are as the following.

\mybox{Context: Find the intersection of two input. You can check every element in set1 one by one.}

\mybox{(0)=LLM("Find the intersection for [\{(Set1)\}[0]] and \{(Set2)\}. Output [] if mutually exclusive.") \\
(1)=LLM("Find the intersection for [\{(Set1)\}[1]] and \{(Set2)\}. Output [] if mutually exclusive.") \\
... \\
(length-1)=LLM("Find the intersection for [\{(Set1)\}[length-1]] and \{(Set2)\}. Output [] if mutually exclusive.") \\
(length)=LLM("Combine {(0)}, {(1)}, {(2)},  ... ,\{(length-1)\} in one array.")}

\paragraph{Arithmetic Calculation} The context description and the \script-script example are as the following.

\mybox{Context: Perform the arithmetic result of input .You can only operate two numbers at a time. Calculate from left to right. Do multiplication and division first.}

\mybox{(0)=LLM("Given \{(input)\}, Split the numbers without operators. Only output list.") \\
(1)=LLM("Add(\{(0)\}[0], \{(0)\}[2]). Only output number. If contains floating point, round to two decimal places.") \\
(2)=LLM("Subtraction(\{(0)\}[1], \{(1)\}). Only output number. If contains floating point, round to two decimal places.") \\
(3)=LLM("Multiply(\{(0)\}[1], \{(1)\}). Only output number. If contains floating point, round to two decimal places.") \\
(4)=LLM("Divide(\{(0)\}[1], \{(1)\}). Only output number. If contains floating point, round to two decimal places.")}

\paragraph{Large Number Addition} The context description and the \script-script example are as the following.

\mybox{Context: Calculate the result of the input. You can plus one digit from one digit starting from the least significant digit.}

\mybox{(0)=LLM("Split "\{(input)\}" by + and output in string format in an array.")\\
(1)=LLM("Calculate \{(0)\}[0][15]+\{(0)\}[1][15]. Only output result.") \\
(2)=LLM("Calculate \{(1)\} divide 10, Only output integer.") \\
(3)=LLM("Calculate \{(2)\}+\{(0)\}[0][14]+\{(0)\}[1][14]. Only output result.") \\
(4)=LLM("Calculate \{(3)\} divide 10, Only output integer.") \\
(5)=LLM("Calculate \{(4)\}+\{(0)\}[0][13]+\{(0)\}[1][13]. Only output result.") \\
(6)=LLM("Calculate \{(5)\} divide 10, Only output integer.") \\
...... \\
(2*length-1)=LLM("Calculate \{(2*length-2)\}+\{(0)\}[0][0]+\{(0)\}[1][0]. Only output result.") \\
(2*length)=LLM("Calculate \{(2*length-1)\} divide 10, Only output integer.") \\
(2*length+1)=LLM("Convert into an integer: \{(2*length)\}\{(2*length-1)\}[-1]\{(2*length-3)\}[-1]\{(2*length-5)\}[-1]......\{(7)\}[-1]\{(5)\}[-1]\{(3)\}[-1]\{(1)\}[-1]")}

\section{Experiment Details}
\label{app:experiment_details}
In the following, we provide experiment details, including the task query example and manual prompt designs utilized for the baseline methods.

\subsection{Task Query Example}
\label{appsub:task_query_example}

In this section, we will provide one input query example for each task.

\paragraph{Yelp Review Comprehension} The input is an array contains ten reviews. We only list one of them in the following:

\mybox{\textit{For starters the food sucks and the staff  at night are very rude. If you don't know what you want right away they skip over you and have no patience. Glass is broken all the time. People get into fights. The bouncers who kick people out dragged an African American man out by his hood across the floor. They kick people out for just not liking them. They go over maximum capacity all the time. The bar deserves to be shutdown.}}

\paragraph{Keyword Counting} The input is an article contains 14 to 20 sentences. We only list one of them in the following:

\mybox{\textit{John, an avid traveler from Canada, had spent his summer exploring the heart of Australia, specifically, the Outback. The vast, arid landscapes of Australia presented a stark contrast to the snow-filled winters of his home in Canada, and he reveled in the difference. He then shared stories of his trip to Brazil, where he fell in love with the vibrant rhythms and the people's warm hospitality. Indeed, Brazil left such a strong impression on him that he visited the country again, this time to explore the dense Amazon rainforest. As John recounted his travels, his friend Sarah, a history buff from the United Kingdom, couldn't help but gush about her trips to Italy and Greece. She explained how she had spent weeks soaking up the culture, history, and mythology of Italy and Greece. Intrigued by Sarah's stories, John revealed his fascination for Northern countries, particularly Norway and Sweden. He cherished his memories of hiking through the scenic landscapes of Norway and the breathtaking fjords of Sweden. Sarah, not to be outdone, discussed her recent visit to Mexico and Cuba. Highlighting the unique colonial architecture of Mexico and the vibrant music scene in Cuba, Sarah couldn't conceal her wanderlust. She ended the conversation by expressing her desire to visit South Korea and Japan. She was particularly interested in the modern cities and ancient temples of South Korea, as well as the unique blend of tradition and technology in Japan. As they parted ways, both agreed to continue exploring and understanding the world, one country at a time.}}

\paragraph{Sorting} In sorting task, the input is an unsorted array contains duplicate numbers.

The following example is length 16:
\mybox{\textit{[1, 2, 6, 1, 1, 6, 0, 3, 7, 4, 5, 2, 9, 2, 1, 5]}}

The following example is length 32:
\mybox{\textit{[0, 0, 5, 9, 0, 7, 9, 9, 1, 2, 6, 1, 1, 9, 0, 1, 3, 5, 2, 3, 5, 6, 0, 2, 7, 4, 6, 2, 9, 7, 9, 5]}}

The following example is length 64:
\mybox{\textit{[6, 3, 6, 5, 1, 2, 4, 3, 8, 0, 7, 8, 6, 4, 9, 5, 2, 4, 8, 4, 4, 4, 5, 6, 8, 4, 7, 7, 8, 9, 4, 9, 5, 4, 8, 4, 0, 5, 6, 9, 1, 2, 3, 6, 2, 0, 8, 1, 0, 7, 1, 2, 0, 7, 6, 9, 9, 9, 5, 6, 8, 3, 9, 0]}}

\paragraph{Set Operation} The input contains two sets without duplicate numbers.

The following example is length 32:
\mybox{Set1: [11, 60, 1, 49, 21, 33, 14, 56, 54, 15, 23, 40, 45, 22, 7, 28, 20, 46, 51, 6, 34, 37, 3, 50, 17, 8, 25, 0, 35, 47, 18, 19] \\
Set2: [31, 11, 4, 63, 38, 58, 59, 24, 61, 14, 32, 39, 27, 46, 48, 19, 52, 57, 50, 56, 3, 2, 53, 29, 5, 37, 62, 41, 36, 12, 49, 16]}

The following example is length 64:
\mybox{Set1: [42, 73, 86, 39, 85, 77, 69, 59, 43, 127, 121, 88, 109, 53, 70, 66, 25, 51, 34, 78, 45, 11, 40, 99, 68, 47, 49, 41, 101, 31, 24, 84, 36, 29, 118, 75, 3, 27, 30, 80, 125, 8, 37, 46, 90, 21, 60, 83, 19, 6, 95, 117, 87, 18, 100, 13, 22, 10, 110, 102, 35, 81, 17, 63] \\
Set2: [34, 49, 116, 106, 112, 23, 5, 80, 18, 62, 90, 54, 32, 103, 37, 43, 9, 25, 92, 16, 111, 79, 64, 91, 107, 58, 72, 94, 7, 60, 33, 14, 19, 104, 28, 74, 96, 76, 38, 52, 114, 50, 17, 0, 3, 100, 69, 98, 2, 1, 99, 12, 95, 97, 123, 4, 126, 124, 82, 27, 67, 57, 115, 46]}

The following example is length 128:
\mybox{Set1: [132, 75, 157, 25, 199, 202, 147, 109, 221, 110, 220, 251, 213, 11, 224, 101, 200, 170, 155, 71, 119, 122, 39, 1, 29, 113, 189, 212, 10, 219, 49, 28, 151, 40, 103, 8, 145, 214, 114, 91, 175, 107, 152, 163, 148, 246, 176, 181, 18, 106, 74, 115, 144, 0, 205, 121, 46, 234, 142, 223, 228, 162, 96, 97, 130, 156, 172, 241, 33, 186, 137, 150, 65, 161, 226, 116, 111, 12, 146, 38, 167, 4, 108, 169, 61, 93, 190, 252, 22, 31, 3, 9, 13, 35, 23, 141, 129, 198, 85, 84, 62, 158, 201, 67, 117, 59, 41, 191, 56, 90, 51, 227, 143, 83, 184, 174, 125, 98, 232, 238, 57, 225, 54, 179, 177, 237, 37, 95] \\
Set2: [27, 162, 187, 254, 128, 227, 2, 165, 143, 109, 140, 46, 160, 26, 139, 171, 42, 199, 207, 30, 205, 117, 213, 48, 40, 212, 185, 196, 197, 94, 136, 35, 229, 193, 36, 7, 15, 43, 4, 203, 142, 144, 49, 31, 240, 124, 116, 69, 37, 250, 95, 105, 103, 168, 126, 64, 73, 206, 24, 157, 135, 118, 34, 134, 45, 62, 153, 5, 47, 239, 216, 222, 80, 231, 102, 21, 57, 215, 149, 141, 236, 32, 188, 204, 194, 23, 233, 83, 154, 210, 159, 70, 202, 253, 20, 71, 166, 242, 221, 228, 78, 230, 29, 145, 147, 81, 104, 235, 66, 100, 131, 132, 244, 195, 68, 72, 53, 182, 79, 248, 3, 82, 211, 173, 180, 17, 77, 51]}

\paragraph{Arithmetic Calculation} The input contains an arithmetic sequence.

The following example is length 8:
\mybox{\textit{6+4+3+3*3+2+4+2}}

The following example is length 16:
\mybox{\textit{2/9-3-4+6+4-9+8+8-4*5-7+2/1+6+7}}

The following example is length 32:
\mybox{\textit{8-2/2/9+9*1/7/3*4+2/5-9+4*8+5+8+9+5+5-2+7/2-2+6-8+7+6+5+1+6*3+1}}

\paragraph{Large Digit Addition} The input is the addition of two large digit numbers.

The following example is length 8:
\mybox{\textit{57247728+67594862}}

The following example is length 16:
\mybox{\textit{5465458164972518+8654164596886757}}

The following example is length 32:
\mybox{\textit{59842829133617473427166884252972+\\24873376371863371698982744892145}}

\subsection{Manual Prompts Design for Baseline Prompt Schemes}
\label{appsub:manual_prompt_baseline}

We provide manual designs for task not covered in baseline prompt schemes. In particular, follows the prompt template for \textbf{keyword counting}, \textbf{sorting}, and \textbf{set intersection} for CoT, ToT, and GoT based on the open source code provided by GoT~\cite{besta2024graph}. The following details prompt design for the \textbf{arithmetic calculation} task. Please view the detail prompt for \textbf{Yelp review comprehension} and \textbf{large digit addition} in our anonymized Github repository \github.

\paragraph{Few shot arithmetic example for Chain-of-thoughts} We provide an step-by-step calculation example for CoT prompt scheme as follows. In particular, we present the an example of the same problem size as the target task. It is worth noting that \method leverages the same example of \textit{shorter} problem size across different problem size, demonstrating its effectiveness in reducing human labor.

\mybox{\tagger{Example} \\
Input: 3+5+6+2+4+5*3+2 \\
Answer: 3+5=8, 8+6=14, 14+2=16, 16+4=20, 5*3=15, 20+15=35, 25+2=37. \\The final answer is 37.
}

\mybox{
\tagger{Example} \\
Input: 7+4+1*6+7+3+7+2+2*7+3+3*6+2+5+4 \\
Answer: 7+4=11, 1*6=6, 11+6=17, 17+7=24, 24+3=27, 27+7=24, 34+2=36, 2*7=14, 36+14=50, 50+3=53, 3*6=18, 53+18=71, 71+2=73, 73+5=78, 78+4=82.\\The final answer is 82.
}
\mybox{
\tagger{Example} \\
Input: 7+6+2+7+3+6+5*2+4+2+4+7+2+4+3*3\\+3+5+4+7+6+4+6+7+6+5*2*7+7+3+7+7 \\
Answer: 7+6=13, 13+2=15, 15+7=22, 22+3=25, 25+6=31, 52=10, 31+10=41, 41+4=45, 45+2=47, 47+4=51, 51+7=58, 58+2=60, 60+4=64, 33=9, 64+9=73, 73+3=76, 76+5=81, 81+4=85, 85+7=92, 92+6=98, 98+4=102, 102+6=108, 108+7=115, 115+6=121, 527=70, 121+70=191, 191+7=198, 198+3=201, 201+7=208, 208+7=215. \\The final answer is 215.
}

\paragraph{Manual Module Preparation for ToT} 
The detail mechanism for ToT and GoT is introduced as follows. ToT first repeatedly use a \textit{Calculation Module} to perform initial calculation attempts. Then, it selects the best attempt by prompting the LLM to score its own attempt. Afterwards, ToT leverages the \textit{Improve Module} to make improvements based on the current and previous results (i.e., a node and its parent node in the tree of thoughts).

Following~\cite{besta2024graph}, the \textit{Calculation Module} includes few shot examples for the targeted problem sizes and is prepared as follows.

\mybox{
\tagger{Instruction} Calculate the given sequence. Output only number, no additional text.\\
\tagger{Example} \\
Input: 3+5+6+2+4+5*3+2 \\
Output: 37 \\
Input: 7+4+7*6+7+3+7+2+7*7+3+3*6+2+5+4 \\
Output: 153 \\
Input: 7+6+2+7+3+6+5*2+4+2+4+7+2+4+3*3+3+\\5+4+7+6+4+6+7+6+5*2*7+7+3+7+7 \\
Output: 215
}

The \textit{Improve Module} is prepared as follows. 
\mybox{\tagger{Instruction} There are some errors in the following calculation sequence. Fine the errors in it and correct it.\\
\tagger{Approach} \\
To fix the incorrectly answer follow these steps: \\
1. Check all numbers in the sequence one by one. \\
2. Attention to the symbol error using.\\
\tagger{Example} \\
Input: 3+5+6+2+4+5*3+2 \\
Incorrectly Answer: 39 \\
Reason: Add 2 one more time \\
Output: 37 \\
Input: 7+4+7*6+7+3+7+2+7*7+3+3*6+2+5+4 \\
Incorrectly Answer: 149 \\
Reason: Forget to add 4, the last number in the sequence \\
Output: 153 \\
Input: 7+6+2+7+3+6+5*2+4+2+4+7+2+4+3*3\\+3+5+4+7+6+4+6+7+6+5*2*7+7+3+7+7\\
Incorrectly Answer: 202 \\
Reason: The incorrectly addition of the first two numbers, remember to add. \\
Output: 215
}
Specifically, we record the mistakes that LLMs make under the ToT scheme, and iteratively append the corrections into the \textit{Improve Module}.

\paragraph{Manual Module Preparation for GoT} 
GoT follows ToT to use the same \textit{Calculation Module} and \textit{Improvement Module}. Besides these, it additionally leverages the \textit{Split Module} and the \textit{Merge Module}. Concretely, GoT first splits the input task into several equal sized chunks, than apply the initial calculation and repeated improvements to each chunk, and finally merge the results from all chunks.

The following is the \textit{Split Module}.

\mybox{\tagger{Instruction} Split the following sequence of 8 numbers into 2 sequence of 4 numbers each, the first sequence should contain the first 4 numbers and the second sequence the second 4 numbers. \\
Only output the final 2 sequence in the following format without any additional text or thoughts!: \\
    ``sequence 1": 3+4+5*1+, \\
    ``sequence 2": 5+2+3*4 \\
\tagger{Example} \\
Input: 3+5+6+2+4+5*3+2 \\
Output:  \\
    ``sequence 1": 3+5+6+2+, \\
    ``sequence 2": 4+5*3+2 \\
Input: 7+4+7*6+7+3+7+2+7*7+3+3*6+2+5+4 \\
Output:  \\
    ``sequence 1": 7+4+7*6+7+3+7+2+, \\
    ``sequence 2": 7*7+3+3*6+2+5+4 \\}

The following is the \textit{Merge Module}. 

\mybox{
\tagger{Instruction} Merge the following 2 final answers.
Only output the final number without any additional text or thoughts!\\
\tagger{Approach} \\
To merge the two number, follow these steps: \\
1. Calculate the answer of 2 numbers \\
\tagger{Example} \\
Input:  \\
    ``sequence 1": 14, \\
    ``sequence 2": 28 \\
Output: 42}

It is worth noting that the split-then-merge structure is antithetical to the arithmetic calculation task, as the operation in the middle of the sequence is not necessarily addition or subtraction. Nevertheless, we design the addition operation for the merge module and further evaluate all prompt schemes with pure addition sequences. As shown in Fig.~\ref{fig:add} and Table~\ref{tab:additional_arithmetic}, finds better performances compared to arithmetic calculations involving all four operations. However, it still fails in comparison to \method as well as CoT-based schemes.

\end{document}